\documentclass[twocolumn]{aastex631}

\DeclareRobustCommand{\textsc}[1]{{\mdseries\scshape #1}}
\providecommand{\uat}[2]{\href{http://astrothesaurus.org/uat/#2}{#1 (#2)}}
\makeatletter
\@ifundefined{@keys@name}{}{%
  \def\@keys@name{\textit{Unified Astronomy Thesaurus concepts:}\/~\mbox{}}}
\@ifundefined{frontmatter@keys@format}{}{%
  \g@addto@macro\frontmatter@keys@format{\hangindent=0pt\hangafter=0\relax}}
\makeatother

\makeatletter
\def\sec@upcase#1{#1}
\def\abstractname{Abstract}
\long\def\@makecaption#1#2{\vskip\abovecaptionskip{\normalsize{\bf #1.} #2\par}\vskip\belowcaptionskip}
\def\fnum@figure{Figure~\thefigure}
\def\fnum@table{Table~\thetable}
\makeatother
\usepackage{etoolbox}
\patchcmd{\thebibliography}{REFERENCES}{References}{}{\PackageWarning{main}{References heading patch failed}}

\usepackage{amsmath,amstext,mathrsfs}
\usepackage{bm}

\defcitealias{LazarianPogosyan2016}{LP16}
\defcitealias{Lazarian2026}{LPM26}
\defcitealias{Melnichenka2026}{MLP26}
\newcommand{\LP}{\citetalias{LazarianPogosyan2016}}
\newcommand{\LPM}{\citetalias{Lazarian2026}}
\newcommand{\MLP}{\citetalias{Melnichenka2026}}
\newcommand{\Lb}{L_{\bar\phi}}
\newcommand{\Gu}{\mathcal{G}_u}
\newcommand{\avg}[1]{\left\langle #1\right\rangle}

\graphicspath{{./}{fig/}}

\shorttitle{Rotation Measure Spectrum from Normalized Polarization Gradients}
\shortauthors{Melnichenka, Lazarian, \& Pogosyan}

\begin{document}

\title{Normalized Polarization Gradients and Other Measures of the Rotation Measure Spectrum}

\author[0009-0004-1969-779X]{A. Melnichenka}
\affiliation{Department of Physics, Berea College, Berea, KY 40404, USA}

\author[0000-0002-7336-6674]{A. Lazarian}
\affiliation{Department of Astronomy, University of Wisconsin--Madison, Madison, WI 53706, USA}

\author[0000-0002-7998-6823]{D. Pogosyan}
\affiliation{Department of Physics, University of Alberta, Edmonton, AB T6G 2E1, Canada}

\begin{abstract}

The spectrum of the Faraday rotation measure traces the turbulence of the magnetized interstellar plasma, but in much of the diffuse Galactic emission the synchrotron emission and Faraday rotation regions coincide, and a single map mixes their fluctuations. Using synthetic observations of \textsc{AthenaK} MHD simulations and synthetic data cubes, we compare on the same maps, for the first time, the normalized tangential polarization gradient (Lazarian et al. 2026) with four other measures, the directional spectrum (Melnichenka et al. 2026a), the spectrum of the polarization and that of the polarization derivative (Lazarian \& Pogosyan 2016), and the spectrum of the polarization gradient (Gaensler et al. 2011 and Zhang \& Liu 2025). The normalized gradient, like the directional spectrum, and the polarization spectrum reach the slope of the rotation measure spectrum at the same wavelength. In the MHD simulations, where the polarized intensity of the source fluctuates, the normalized gradient approaches it faster at weak rotation, and for the strongly polarized synthetic source the two approach it alike. The polarization gradient spectrum traces the emission in the MHD simulations, and the polarization derivative is limited by noise when formed from two nearby wavelengths. Synthetic data cubes confirm the predicted gradient statistics under weak, strong turbulent and strong mean Faraday rotation, with up to three times less gradient power than Gaussian maps under strong turbulent rotation. The results guide single-frequency analyses of LOFAR and SKA data.

\end{abstract}


\section{Introduction}
\label{sec:intro}

Synchrotron polarization carries the statistics of the magnetic field of the emitting plasma, and in much of the diffuse radio sky the same plasma Faraday-rotates the polarization, so that radiation from different depths reaches the observer with different rotation angles \citep{Burn1966, Sokoloff1998}. \citet[][hereafter \LP]{LazarianPogosyan2016} related the correlations of the complex polarization $P=Q+iU$ of such a volume to the spectra of the magnetic field and of the rotation measure (RM) density, and proposed two families of measures, the polarization spatial analysis (PSA) of $P$ and of the polarization derivative $\partial P/\partial\lambda^2$ at a fixed wavelength, and the polarization frequency analysis (PFA) of the polarization variance against wavelength. Their predictions were tested with synthetic observations by \citet{Lee2016}, \citet{Zhang2016} and \citet{Zhang2018}, several synchrotron statistics have since been compared in MHD simulations \citep{Zhang2023}, and the PFA has been applied to surveys \citep{Xiao2026}. These statistics complement rotation measure synthesis, which separates the emission along the line of sight by its Faraday RM but needs a wide coverage in $\lambda^2$ \citep{Brentjens2005}, and the RM structure functions of extragalactic sources \citep{MinterSpangler1996}. The wavelength dependence of the depolarization depends only weakly on the RM spectrum \citep{Tribble1991}.

A separate line of work uses the spatial derivatives of $P$. \citet{Gaensler2011} introduced the magnitude of the polarization gradient $|\nabla P|$, whose map shows filaments associated with jumps of the electron density and of the line-of-sight field. Its moments depend on the sonic Mach number of the turbulence \citep{Burkhart2012}, it has been mapped over large parts of the sky \citep{Iacobelli2014, Herron2017} and on a range of angular scales \citep{Robitaille2015}, and for a uniform background source $|\nabla P|/|P|$ traces the gradient of the RM \citep{Burkhart2012, Iacobelli2014}. \citet{Herron2018a} split the directional derivative of $P$ into a radial part, which changes the polarized intensity, and a tangential part, which turns the polarization angle and is weighted by the polarized intensity, and \citet{Herron2018b} found with MHD simulations and survey data that the tangential part grows with wavelength as the Faraday rotation does. Gradients of the synchrotron intensity and polarization also trace the direction of the magnetic field \citep{Lazarian2017, LazarianYuen2018, Zhang2019}, and \citet{ZhangLiu2025} proposed the power spectrum of the map of $|\nabla P|$ as a measure of the turbulence spectrum under Faraday depolarization.

Per unit polarized intensity $|P|$, the tangential derivative is the gradient of the polarization direction $u=P/|P|$. In \citet[][hereafter \MLP]{Melnichenka2026} we introduced the directional statistics of $u$, among them the directional spectrum, for a Faraday screen in front of the emitting region, and compared them with the measures of \LP. \citet[][hereafter \LPM]{Lazarian2026} normalized the tangential derivative of \citet{Herron2018a} in this way, related its spectrum to the directional spectrum, and derived its statistics for a volume that both emits and rotates, under weak, strong turbulent and strong mean Faraday rotation. A companion paper \citep{Melnichenka2026b} tests the directional correlation and spectrum in this mixed volume.

\begin{figure*}[tp]
\centering
\includegraphics[width=0.90\textwidth]{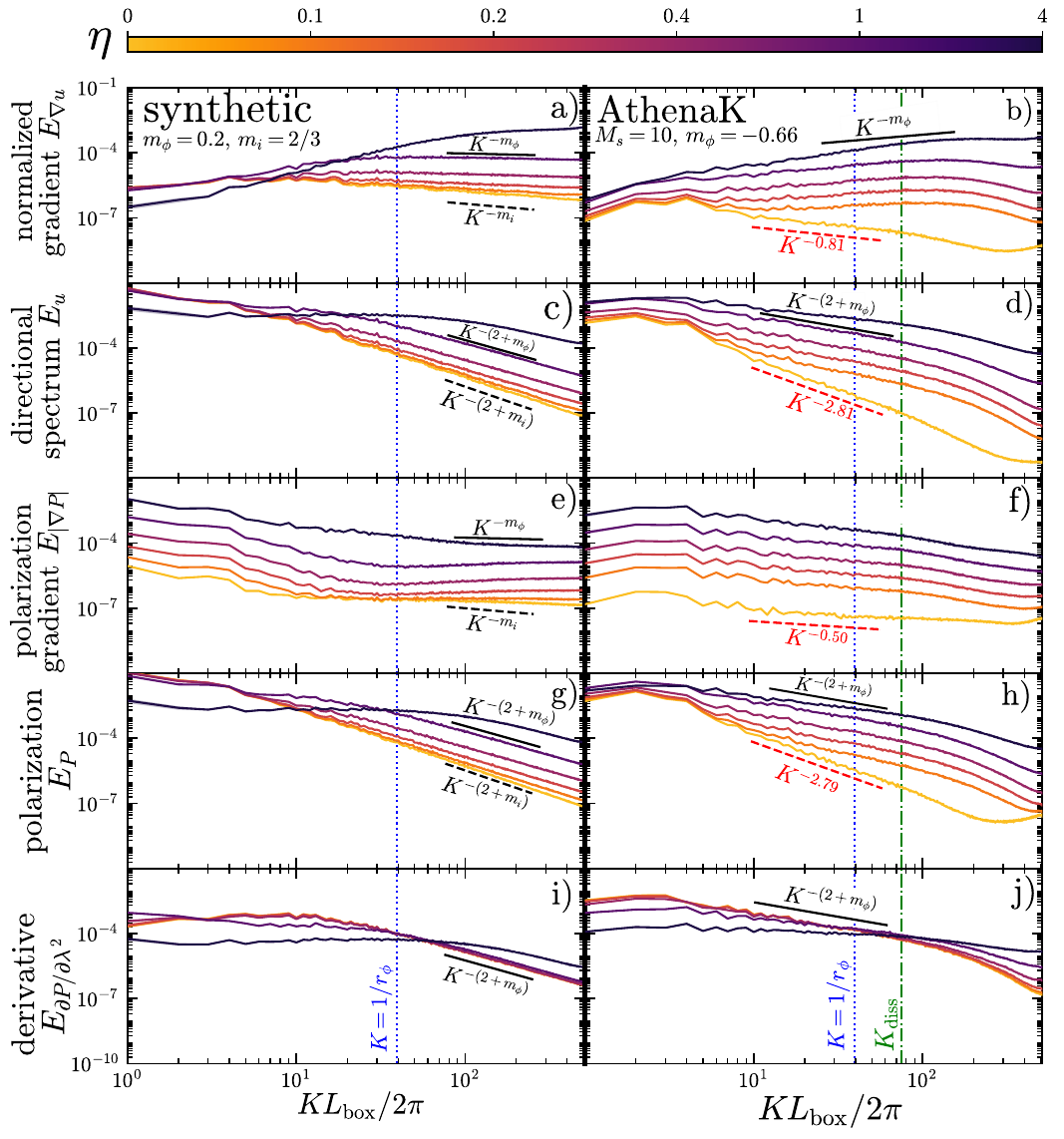}
\caption{Comparison of five polarization measures on the same maps of the \textsc{AthenaK} cube with the sonic Mach number $M_s=10$ (right) and of the synthetic cube (left) at six rotation strengths $\eta$ (Section~\ref{subsec:weak}), from $\eta=0$ without Faraday rotation (yellow) through the transition to strong rotation at $\eta=4$ (dark purple), shown by the color bar. Panels (a, b) show the normalized tangential gradient $\nabla u$ (\LPM), (c, d) the directional spectrum of $u=P/|P|$ (\MLP), (e, f) $|\nabla P|$ \citep{Gaensler2011, ZhangLiu2025}, (g, h) the complex polarization $P$ and (i, j) the polarization derivative $\partial P/\partial\lambda^2$ (\LP). Lines are ring-integrated 1D spectra $E$, the power spectrum times $K$, against $KL_{\rm box}/2\pi$, the wavenumber in units of the fundamental mode of the map of side $L_{\rm box}$. Shaded bands are standard deviations of the mean over observer positions along the line of sight (Section~\ref{subsec:synobs}). Black dashed and solid lines are the predicted intrinsic and Faraday slopes, shifted vertically. In the synthetic cube the Faraday spectrum includes the injection scale of the RM density. In the \textsc{AthenaK} cube $m_\phi<0$, so the Faraday slope of the gradients, $K^{-m_\phi}$, rises with $K$ (Section~\ref{subsec:weak}). The spectrum of $|\nabla P|$ in panel (f) does not show this rise, and the slope is not drawn there. Red dashed lines are power laws fitted at $\eta=0$ in the MHD cube, which has no prescribed index $m_i$ of the emissivity. Blue dotted lines mark $K=1/r_\phi$, the green dash-dotted line $K_{\rm diss}$ the onset of numerical dissipation, which ends the inertial range of the MHD cube.}
\label{fig:compare}
\end{figure*}

\begin{figure}[!htb]
\centering
\includegraphics[width=\columnwidth]{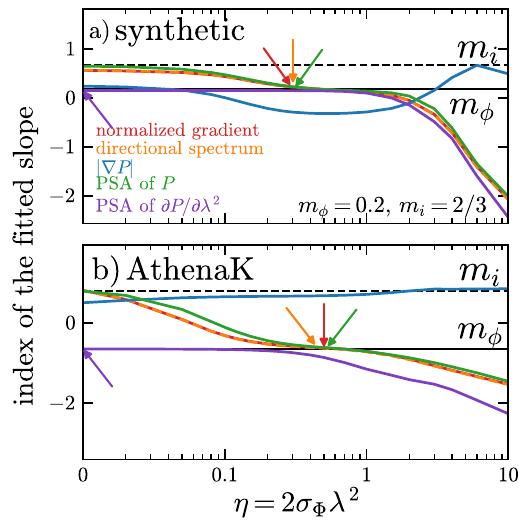}
\caption{Scaling slope of the spectra of Figure~\ref{fig:compare}, fitted over $KL_{\rm box}/2\pi=10$ to 60 in the \textsc{AthenaK} cube, below $K_{\rm diss}$, and 78 to 256 in the synthetic cube, against the rotation strength, in the \textsc{AthenaK} (bottom) and synthetic (top) cubes, written as an index, minus the slope minus 2 for $u$, $P$ and $\partial P/\partial\lambda^2$, and minus the slope for the gradients, so that the predicted Faraday slope is at $m_\phi$ (solid lines) and the intrinsic slope at $m_i$ (dashed lines). In the MHD cube the dashed line is the index of the $\eta=0$ spectrum of $P$. Arrows mark the first $\eta$ at which each measure reaches the Faraday slope. The directional spectrum (dashed orange) lies on the normalized gradient. PSA denotes the spectra of $P$ and $\partial P/\partial\lambda^2$ of the polarization spatial analysis (Section~\ref{sec:intro}).}
\label{fig:compareslope}
\end{figure}

These measures have not been compared on the same maps of a volume in which synchrotron emission and Faraday rotation are spatially coincident. The main result of this paper is such a comparison (Figures~\ref{fig:compare} and \ref{fig:compareslope}). On synthetic observations of \textsc{AthenaK} MHD simulations and of synthetic data cubes we examine which of the normalized tangential gradient, the directional spectrum, the spectra of $P$ and of $\partial P/\partial\lambda^2$, and the spectrum of $|\nabla P|$ recovers the spectrum of the RM density, and under which conditions. We also test the predictions of \LPM{} for the normalized gradient in the three regimes of rotation.

\section{Measures and Predictions}
\label{sec:theory}

\begin{table}[t]
\centering\normalsize
\caption{Notation}
\label{tab:symbols}
\renewcommand{\arraystretch}{1.1}
\begin{tabular}{ll}
\hline\hline
Symbol & Meaning \\
\hline
$P$, $u=e^{2i\chi}$ & Polarization and its direction \\
$\Gu$ & Normalized tangential gradient \\
$K$, $R$ & Wavenumber, separation on the sky \\
$P_u$, $P_{\nabla u}$ & Power spectra of $u$ and of $\nabla u$ \\
$E_u=KP_u$ & Ring-integrated spectra \\
$b$ & Rms width of the Gaussian beam \\
$\phi$, $\Phi$ & RM density and Faraday RM \\
$\bar\phi$, $\sigma_\phi$ & Mean and rms of $\phi$ \\
$r_\phi$, $m_\phi$ & Correlation length, index of $\phi$ \\
$m_i$ & Index of the emissivity $P_i$ \\
$L$ & Depth of the volume \\
$\eta=2\lambda^2\sigma_\Phi$ & Rotation strength \\
$\chi_0$, $q$, $s$ & Coefficients of $\chi$ in powers of $\lambda^2$ \\
$\ell$, $\Lb$ & Distances of one radian of rotation \\
$\rho_P$, $\xi_u$ & Coherence of $P$, correlation of $u$ \\
$\varepsilon=1-\rho_P^2$ & Decorrelation of $P$ \\
$R_c$, $R_a$ & Radii of Equations~\eqref{eq:Rc} and \eqref{eq:Ra} \\
\hline
\end{tabular}
\end{table}

\subsection{The Measures}
\label{subsec:definition}

The largest tangential component of the derivative of the polarization vector $(Q,U)$ along a direction on the sky is $2|P||\nabla\chi|$, where $\chi$ is the polarization angle \citep[][their Equation~(29)]{Herron2018a}, and the notation is collected in Table~\ref{tab:symbols}. Per unit polarized intensity it is the normalized tangential gradient
\begin{equation}
\Gu=|\nabla u|=2|\nabla\chi|
\label{eq:Gu}
\end{equation}
(\LPM, Equation~(34)), the gradient of the direction $u=P/|P|=e^{2i\chi}$. The power spectrum of the vector $\nabla u$ is exactly
\begin{equation}
P_{\nabla u}(K)=K^2P_u(K)
\label{eq:Pgrad}
\end{equation}
(\LPM, Equation~(35)), where $K$ is the wavenumber on the sky and $P_u$ is the directional spectrum of \MLP, so the two spectra carry the same information, while the spectrum of the scalar $\Gu$ is a different, nonlinear statistic. When the map of $u$ is smoothed to a finite angular resolution with a Gaussian beam of rms width $b$, the mean squared gradient of the smoothed map $u_b$ is
\begin{equation}
\avg{|\nabla u_b|^2}=\frac{1}{2\pi}\int_0^\infty dK\,K^3P_u(K)\,e^{-b^2K^2}
\label{eq:beam}
\end{equation}
(\LPM, Equation~(40)).

We compare the gradient with the spectra of $P$ and of $\partial P/\partial\lambda^2$ at a fixed wavelength \citepalias{LazarianPogosyan2016} and with the spectrum of the magnitude of the polarization gradient
\begin{equation}
|\nabla P|=\left[\left(\partial_xQ\right)^2+\left(\partial_xU\right)^2+\left(\partial_yQ\right)^2+\left(\partial_yU\right)^2\right]^{1/2}
\label{eq:zl}
\end{equation}
\citep{Gaensler2011, ZhangLiu2025}, where $x$ and $y$ are the coordinates on the sky. The magnitude $|\nabla P|$ is invariant under translations and rotations in the $Q$--$U$ plane \citep{Gaensler2011}, while $\Gu$ is invariant under rotations only. A uniform Faraday rotation in front of the source leaves it unchanged, and a uniform polarized component added to $Q$ and $U$ changes it.

\subsection{Weak Faraday Rotation}
\label{subsec:weak}

Every element of the volume emits with the complex emissivity $P_i$ and rotates the radiation that passes through it, and the polarization observed at the sky position $\bm X$ and wavelength $\lambda$ from a volume of depth $L$ is
\begin{equation}
P(\bm X,\lambda^2)=\int_0^L dz\,P_i(\bm X,z)\,e^{2i\lambda^2\Phi(\bm X,z)}
\label{eq:Pmixed}
\end{equation}
(\LP{} and \LPM, Equation~(2)), where $\Phi(\bm X,z)$ is the Faraday RM between the observer and the depth $z$, the integral of the RM density, or Faraday rotation density, $\phi\propto n_eB_\parallel$, the product of the thermal electron density and the magnetic field along the line of sight (LOS). The RM density has the mean $\bar\phi$, the rms $\sigma_\phi$ and the correlation $\sigma_\phi^2[1-(r/r_\phi)^{m_\phi}]$ at separations $r$ below the correlation length $r_\phi$, which corresponds to the three-dimensional spectrum $k^{-3-m_\phi}$ at the wavenumber $k$. The index is thus not the slope of the spectrum. The Kolmogorov value $m_\phi=2/3$ corresponds to $k^{-11/3}$, positive $m_\phi$ describes a spectrum steeper than $k^{-3}$, and negative $m_\phi$ a shallower one, whose correlation decreases as $r^{m_\phi}$ (\LP{} and \MLP). The supersonic \textsc{AthenaK} cube has $m_\phi=-0.66$, the shallow spectrum $k^{-2.34}$, far from Kolmogorov, because its RM density is dominated by the fluctuations of the density, which the shocks of the $M_s=10$ flow concentrate at small scales \citepalias{Melnichenka2026}. The emissivity has in the same way the index $m_i$, with $m_i=2/3$ for a Kolmogorov spectrum. Both are indices of the three-dimensional fields, and the indices of the projected maps used in \MLP{} are larger by one. We label the wavelength by the rotation strength $\eta=2\lambda^2\sigma_\Phi$, where $\sigma_\Phi$ is the rms of the Faraday RM of the whole volume. It is the rms rotation of the phase of $P$, $2\lambda^2\Phi$, accumulated through the whole volume, so at $\eta>1$ the radiation from the far part of the volume is turned by more than a radian, by different angles on different lines of sight, and $\eta$ sets the level of Faraday depolarization.

\begin{table*}[t]
\centering\normalsize
\caption{Simulated Data Cubes}
\label{tab:cubes}
\renewcommand{\arraystretch}{1.1}
\begin{tabular*}{\textwidth}{@{\extracolsep{\fill}}lllllll}
\hline\hline
Cube & Source & Resolution & $m_\phi$ & Depth $L$ & Wavelengths & Used in \\
\hline
\textsc{AthenaK} $M_s=10$ & MHD & $1024^3$ & $-0.66$ & $250\,r_\phi$ & $\eta=0$ to 10 & Figs.~\ref{fig:compare}, \ref{fig:compareslope}, \S\ref{subsec:expansion} \\
\textsc{AthenaK} $M_s=1$ & MHD & $1024^3$ & $0.37$ & $32\,r_\phi$ & $\eta=0.05$ to 10 & \S\ref{subsec:expansion} \\
Synthetic & Turbulent & $1024^3$ & $0.2$ & $250\,r_\phi$ & $\eta=0$ to 10 & Figs.~\ref{fig:compare}, \ref{fig:compareslope} \\
Synthetic & Coherent & $1024^3$ & $2/3$ & $47\,r_\phi$ & $\eta=0.05$ to 10 & \S\ref{subsec:expansion} \\
Synthetic, seven cubes & Uniform & $1024^3$ & 0.4 to 1.9 & 6 to $15\,r_\phi$ & $\eta=0.1$ and 0.4 & Fig.~\ref{fig:weak} \\
Synthetic, long LOS & Uniform & $1024^2\times8192$ & $2/3$ & 6 and $47\,r_\phi$ & $\ell=0.05$ to $0.4\,r_\phi$ & Fig.~\ref{fig:near} \\
Synthetic, long LOS & Uniform & $1024^2\times8192$ & $3/2$ & 8 and $63\,r_\phi$ & $\ell=0.02$ to $0.5\,r_\phi$ & Fig.~\ref{fig:near} \\
Synthetic, core & Uniform & $8192^2\times4096$ & $2/3$ & $2\,r_\phi$ & $\ell=0.1$ and $0.2\,r_\phi$ & Fig.~\ref{fig:core} \\
Synthetic, core & Uniform & $8192^2\times20480$ & $3/2$ & $10\,r_\phi$ & $\ell=0.2\,r_\phi$ & Fig.~\ref{fig:core} \\
Synthetic, constant $\phi$ & Turbulent & $1024^2\times4096$ & & 16 and $31\,\Lb$ & $\lambda_1^2$ and $2\lambda_1^2$ & Fig.~\ref{fig:mean} \\
\hline
\end{tabular*}
\vspace{4pt}

{\raggedright \textit{Note.} In the MHD cubes the LOS is perpendicular to the mean field, and the parameters are those measured in \MLP. For the synthetic cubes $r_\phi$ is the correlation length in the small-separation form of the prescribed correlation, and for the cube used in the comparison, with $m_\phi=0.2$, it is the separation at which the correlation falls to one half.\par}
\end{table*}

When the rotation is small on every line of sight that contributes, the polarization angle is
\begin{equation}
\chi=\chi_0+\lambda^2q+\lambda^4s+\mathcal{O}(\lambda^6)
\label{eq:chi}
\end{equation}
(\LPM, Equation~(A2)), where $\chi_0$ is the angle of the unrotated polarization, $q$ is the Faraday RM weighted by the emission behind each depth, and $s$ is the coefficient of second order. The squared gradient is then
\begin{equation}
\begin{aligned}
\Gu^2=\;&4|\nabla\chi_0|^2+8\lambda^2\,\nabla\chi_0\cdot\nabla q\\
&+4\lambda^4\left[|\nabla q|^2+2\,\nabla\chi_0\cdot\nabla s\right]+\mathcal{O}(\lambda^6)
\end{aligned}
\label{eq:expansion}
\end{equation}
(\LPM, Equation~(36)). The first term is the intrinsic gradient, the term in $\lambda^2$ is a cross-correlation between the intrinsic direction and the Faraday RM, and the term in $\lambda^4$ contains the gradient of the Faraday RM. For a source that is uniform along the line of sight and has one intrinsic direction, only the last term remains, and the gradient spectrum is its Faraday term
\begin{equation}
P_{\nabla u,F}(K)=4\lambda^4K^2P_q(K)\propto\lambda^4K^{-(1+m_\phi)}
\label{eq:weakspec}
\end{equation}
(\LPM, Equation~(37)) for every $0<m_\phi<2$, where $P_q$ is the spectrum of $q$. For $m_\phi>1$ the leading $R^2$ term of the structure function of $q$ at small separations $R$ on the sky carries no power at nonzero wavenumbers and does not set the slope.

The rotation thus moves the scaling slope of each spectrum from that of the emission, the intrinsic slope, to that of the RM density, the Faraday slope. The comparison uses the ring-integrated 1D spectra of \citet{Zhang2018}, the power spectrum times $K$. In the inertial range of $E_u=KP_u$, $E_P$ and $E_{\partial P/\partial\lambda^2}$ the intrinsic and the Faraday terms go as $K^{-(2+m_i)}$ and $K^{-(2+m_\phi)}$ \citepalias{LazarianPogosyan2016, Lazarian2026}, and in $E_{\nabla u}$ as $K^{-m_i}$ and $K^{-m_\phi}$. The derivative $\partial P/\partial\lambda^2$ weights the emission from each depth by its Faraday RM and carries the Faraday slope already at the shortest wavelengths \citepalias{LazarianPogosyan2016}.

\section{Simulations and Synthetic Observations}
\label{sec:numerics}

\subsection{AthenaK MHD Simulations}
\label{subsec:athena}

The MHD cubes (Table~\ref{tab:cubes}) are two snapshots of the isothermal \textsc{AthenaK} \citep{Stone2026} runs of \MLP, with the resolution $1024^3$, the Alfv\'en Mach number $M_A=0.8$, so that the magnetic field $\bm B$ fluctuates around a dynamically important mean field $\bm B_0$, and the sonic Mach numbers $M_s=1$ and 10. Each element emits with $P_i\propto(B_1+iB_2)^2$, where $B_1$ and $B_2$ are the field components in the plane of the sky. This is the synchrotron emissivity of relativistic electrons with a power-law energy spectrum of index 3, uniform in space, for which the polarized intensity is proportional to the square of the magnetic field in the plane of the sky. Each element also rotates with $\phi\propto n_eB_\parallel$, where $B_\parallel$ is the field component along the LOS. Their turbulence is the cascade of compressible MHD turbulence \citep{Goldreich1995, ChoLazarian2003}. The LOS is perpendicular to the mean field, which then gives every line of sight a net polarization, as the expansion of Equation~\eqref{eq:chi} requires.

\subsection{Synthetic Data Cubes}
\label{subsec:synthetic}

In the synthetic data cubes the RM density is a Gaussian random field generated in wavenumber space. It has zero mean, except in the cube with a constant RM density, and the spectrum $P_\phi(k)\propto(k^2+r_0^{-2})^{-(3+m_\phi)/2}$ with the injection scale $r_0$, and the emissivity and the RM density are independent. Most cubes have a uniform source, the case of Equation~\eqref{eq:weakspec} and of the predictions for strong turbulent Faraday rotation (Section~\ref{subsec:near}). For strong turbulent Faraday rotation two of them have a long LOS, to cover neighboring lines of sight from low to almost full correlation, and two others have a wide sky and are repeated with eight realizations of the RM density. The coherent source is constant along each line of sight and has an intrinsic angle that varies across the sky, which gives it an intrinsic gradient. The turbulent source, used for the comparison and for the mean Faraday rotation, is made from a synthetic magnetic field whose components in the plane of the sky are $B_1=1+0.6\,g_1$ and $B_2=0.6\,g_2$, where $g_1$ and $g_2$ are independent Gaussian random fields with the Kolmogorov spectrum, and it emits with $P_i\propto(B_1+iB_2)^2$, as in the MHD cubes and in \MLP. The Gaussian fields of the synthetic cubes are therefore the RM density and the fluctuations of the magnetic field, while this emissivity, quadratic in the magnetic field, is not Gaussian and has a mean polarization, mimicking synchrotron emission. It is strongly polarized, and its polarized intensity varies little across the sky.

\subsection{Synthetic Observations}
\label{subsec:synobs}

Equation~\eqref{eq:Pmixed} is integrated along the LOS, the map of $u$ is formed, and its gradient, like the derivatives of $Q$ and $U$ in Equation~\eqref{eq:zl}, is taken in Fourier space. The cubes are periodic, so the observer can be placed at several positions along the LOS, and statistics are averaged over these positions or over realizations, with the standard deviation of the mean as the error. The predictions contain no free parameters and are computed from the same cube or from the prescribed spectrum of the RM density. As a reference with Gaussian statistics we randomize the Fourier phases of each polarization map, which keeps its power spectrum and two-point correlation. For the comparison of measures $\partial P/\partial\lambda^2$ is computed exactly on noise-free maps. In a separate noise test we add to $Q$ and $U$ Gaussian noise with an rms of $1\%$ of the mean synchrotron intensity, then smooth the maps to the angular resolution of a telescope with a Gaussian beam with a full width at half maximum of three times the spacing of the lines of sight, as in \citet{Zhang2016, Zhang2018}, and form $\partial P/\partial\lambda^2$ from maps at two wavelengths whose values of $\lambda^2$ differ by 6\%, with independent noise.

\section{Comparison of Measures}
\label{sec:compare}

\begin{figure}[t]
\centering
\includegraphics[width=\columnwidth]{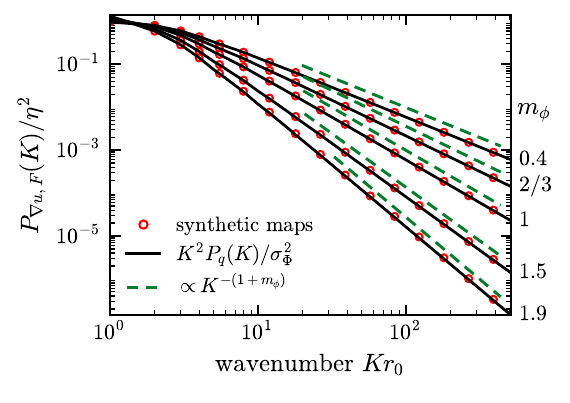}
\caption{Gradient spectrum at weak Faraday rotation, divided by $\eta^2$, for synthetic cubes with a uniform source and $m_\phi=0.4$, 2/3, 1, 1.5 and 1.9 from top to bottom, at $\eta=0.1$ (markers), against $Kr_0$. Black lines are $K^2P_q/\sigma_\Phi^2$ of Equation~\eqref{eq:weakspec}, evaluated from the prescribed spectrum, and green dashed segments the slope $-(1+m_\phi)$, shifted upward.}
\label{fig:weak}
\end{figure}

\begin{figure*}[t]
\centering
\includegraphics[width=0.98\textwidth]{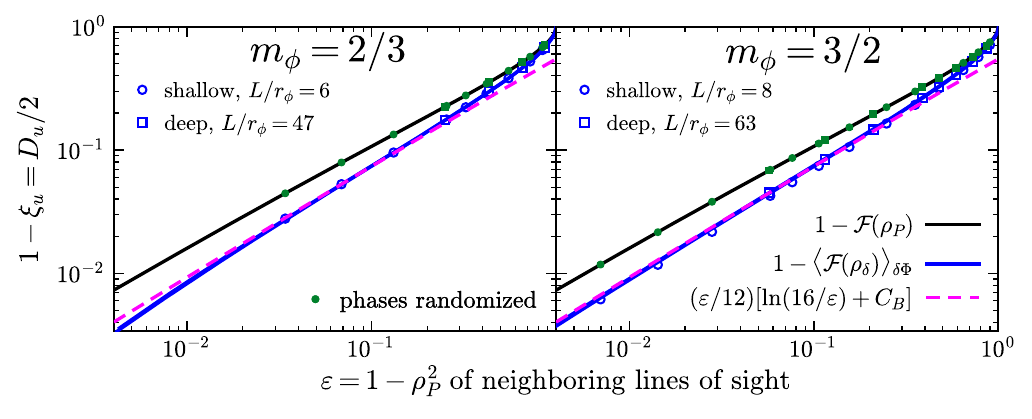}
\caption{Decorrelation of the directions of neighboring lines of sight against the decorrelation of the polarization, for synthetic cubes with a uniform source at a shallow (circles) and a deep (squares) depth, with $L/r_\phi$ given in each panel. Both panels have the same depths in cells, and $L/r_\phi$ differs because $r_\phi$ depends on $m_\phi$. Each open marker is one wavelength, filled markers are the same maps with randomized Fourier phases. The black line is the Gaussian pair, Equation~\eqref{eq:F}, the blue lines the conditional average of Equation~\eqref{eq:cond}, and the dashed line its near-coherence limit for a deep volume, Equation~\eqref{eq:brown}. The vertical axis $1-\xi_u$ is half the structure function $D_u$ of $u$.}
\label{fig:near}
\end{figure*}

The normalized gradient and the directional spectrum are one statistic by Equation~\eqref{eq:Pgrad}, and their slopes coincide at every rotation strength. All three have the slope of the emission without rotation and take the slope of the RM density at the same rotation strength, as Equation~\eqref{eq:weakspec} predicts and as \citet{Lee2016} and \citet{Zhang2018} found for $P$. In the MHD cube the spectrum of $P$ flattens from $K^{-2.79}$ without rotation to the predicted $K^{-1.34}$ at $\eta\approx0.7$, and in the synthetic cube from $K^{-2.65}$ to $K^{-2.2}$ at $\eta\approx0.4$. In both cubes the Faraday spectrum is much shallower than that of the emission, so this transition is clear. In the MHD cube the slope of the normalized gradient approaches the Faraday slope faster at weak rotation. Its index differs from the predicted one by 0.34 and 0.13 at $\eta=0.1$ and 0.2, and that of $P$ by 0.55 and 0.21 (Figure~\ref{fig:compareslope}). We attribute this to the spectra of $u$ and $\nabla u$ responding to the rotation through the polarization angle alone, while the spectrum of $P$ also contains the fluctuations of the polarized intensity of the source, which are present without rotation. For the synthetic source, whose polarized intensity varies little, the two approach the Faraday slope alike. The spectrum of $\partial P/\partial\lambda^2$ carries the Faraday term already without rotation (Section~\ref{subsec:weak}).

Without rotation the slope of the $|\nabla P|$ spectrum differs from that of $K^2E_P$ by 0.3 in the supersonic MHD cube, whereas \citet{ZhangLiu2025} found this relation in their MHD simulations, and in the synthetic cube the relation fails, as they also found for isotropic turbulence. Under rotation its slope in the MHD cube changes by less than 0.4, as \citet{ZhangLiu2025} found, and stays within 0.3 of that of the emission, below the onset of numerical dissipation, where the normalized gradient already takes the Faraday slope, so it traces the emission rather than the RM density. In the synthetic cube its slope passes the Faraday slope at $\eta\approx0.1$, keeps rising to $K^{0.32}$ at $\eta\approx0.5$, and falls again under strong rotation. The spectrum of the vector gradient of $P$, formed from the four derivatives of $Q$ and $U$ before the magnitude is taken, is exactly $K^2E_P$ and takes the Faraday slope like the normalized gradient, $K^{0.70}$ at $\eta=1$ in the MHD cube against the predicted $K^{0.66}$. In that cube the spectrum of the magnitude in Equation~\eqref{eq:zl} does not take it, with central differences as with spectral derivatives, so the difference is not numerical. The spectrum of the scalar $|\nabla u|$ reaches the Faraday slope in the synthetic cube but not in the MHD cube, so in our cubes the vector gradient, Equation~\eqref{eq:Pgrad}, is the reliable form.

In the noise test of Section~\ref{subsec:synobs} the spectrum of $\partial P/\partial\lambda^2$ does not reach the Faraday slope, because the difference of two maps enhances the noise \citep{Zhang2016}. Without noise it has the Faraday slope already at the shortest wavelengths (Section~\ref{subsec:weak}), as \LP{} predicted, and is the most direct measure of the RM spectrum. In the MHD cube the normalized gradient and the spectrum of $P$ still reach it, at a larger rotation strength than without noise, as in \citet{Lee2016} and \citet{Zhang2018}. In the synthetic cube, whose intrinsic and Faraday slopes differ by only 0.47, the beam steepens both spectra within the fitted range, and neither reaches it. At $\eta$ of a few the Faraday depolarization reaches the fitted wavenumbers, and none of the measures keeps the Faraday slope there, as for the spectra of \citet{Lee2016} and \citet{Zhang2018}.

\begin{figure*}[t]
\centering
\includegraphics[width=0.98\textwidth]{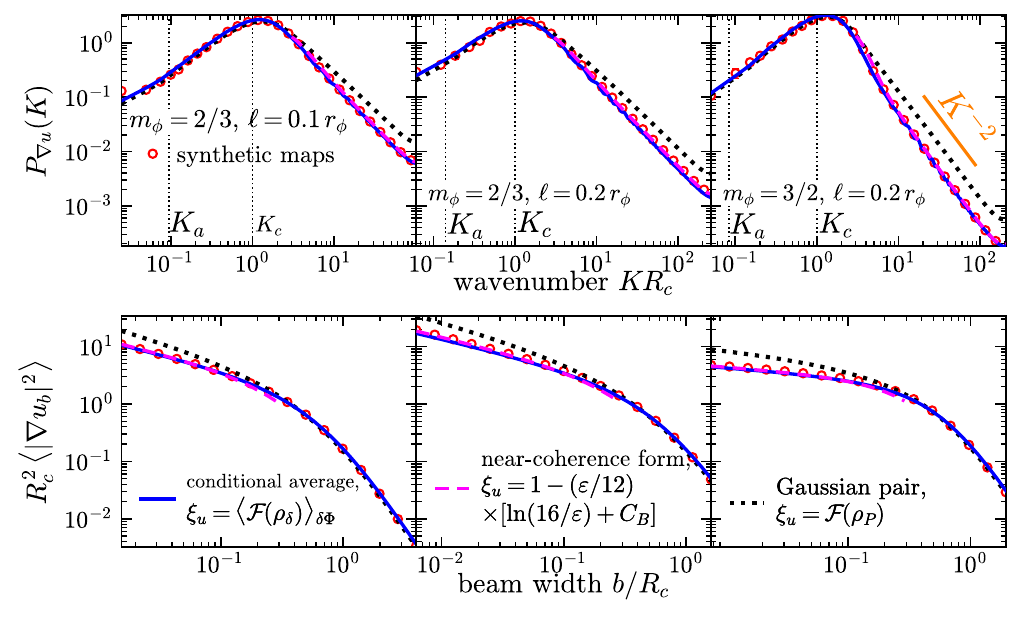}
\caption{Gradient spectrum (top) and beam-smoothed gradient variance (bottom) of synthetic cubes with a uniform source that resolve the coherent core, for two wavelengths at $m_\phi=2/3$ and one at $m_\phi=3/2$. Markers are means over the realizations. The lines are computed with Equations~\eqref{eq:Pgrad} and \eqref{eq:beam} from the conditional average (solid), the near-coherence form (dashed) and the Gaussian pair (dotted), each with the coherence of \LP. Vertical lines mark $K_a$ and $K_c$, and the segment labeled $K^{-2}$ marks that slope, shifted upward.}
\label{fig:core}
\end{figure*}

\begin{figure*}[t]
\centering
\includegraphics[width=0.98\textwidth]{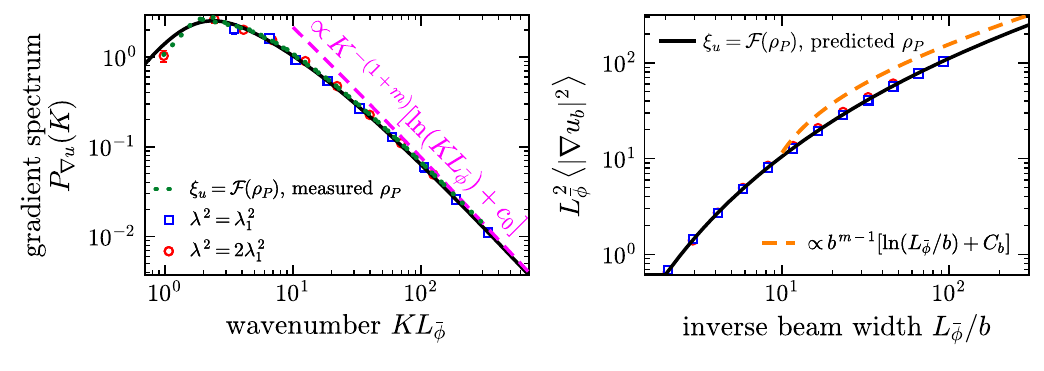}
\caption{Constant RM density with the synthetic turbulent source, $m_i=2/3$, at two wavelengths, $\lambda^2=\lambda_1^2$ and $2\lambda_1^2$. The left panel shows the gradient spectrum against $K\Lb$, the right panel the beam-smoothed gradient variance in units of $\Lb^{-2}$ against $\Lb/b$. Markers are measured. Solid lines are computed from the Gaussian pair, Equation~\eqref{eq:F}, with the coherence of Equation~\eqref{eq:meanrho}, and the dotted line with the coherence measured on the same maps. The dashed lines are the asymptotic forms of Equations~\eqref{eq:meanP} and \eqref{eq:meanV}, labeled with $m=m_i$.}
\label{fig:mean}
\end{figure*}

\begin{table}[!htb]
\centering
\caption{Which Measure Should Be Preferred in Our Tests}
\label{tab:winner}
\renewcommand{\arraystretch}{1.1}
\begin{tabular}{lll}
\hline\hline
Data & Preferred & Reason \\
\hline
\parbox[t]{0.28\columnwidth}{\raggedright Single frequency, fluctuating polarized intensity} & \parbox[t]{0.25\columnwidth}{\raggedright normalized gradient (or $P_u$)} & \parbox[t]{0.34\columnwidth}{\raggedright same onset as $P$, nearer the Faraday slope at weak rotation} \\[2pt]
\parbox[t]{0.28\columnwidth}{\raggedright Single frequency, strongly polarized source} & \parbox[t]{0.25\columnwidth}{\raggedright normalized gradient or $P$} & \parbox[t]{0.34\columnwidth}{\raggedright same onset and response} \\[2pt]
\parbox[t]{0.28\columnwidth}{\raggedright Two wavelengths, no noise} & \parbox[t]{0.25\columnwidth}{\raggedright $\partial P/\partial\lambda^2$} & \parbox[t]{0.34\columnwidth}{\raggedright Faraday slope from $\eta=0$} \\[2pt]
\parbox[t]{0.28\columnwidth}{\raggedright Two wavelengths, with noise} & \parbox[t]{0.25\columnwidth}{\raggedright normalized gradient or $P$} & \parbox[t]{0.34\columnwidth}{\raggedright derivative limited by noise} \\[2pt]
\parbox[t]{0.28\columnwidth}{\raggedright Any} & \parbox[t]{0.25\columnwidth}{\raggedright not $|\nabla P|$} & \parbox[t]{0.34\columnwidth}{\raggedright does not keep the Faraday slope} \\[2pt]
\hline
\end{tabular}
\end{table}

In our tests the choice of measure for the spectrum of the RM density therefore depends on the data (Table~\ref{tab:winner}). At a single frequency the normalized gradient, or equivalently the directional spectrum, has advantages over the spectrum of $P$ when the polarized intensity of the source fluctuates, as in the MHD cube, and performs as well as it for a strongly polarized source. With noise-free maps at two wavelengths the polarization derivative gives the most direct measure, while at the noise of our test the single-frequency measures are more robust. The spectrum of the magnitude of the polarization gradient is better suited to the emission, the use proposed by \citet{ZhangLiu2025}, than to the RM density.

\section{Tests of the Gradient Predictions}
\label{sec:tests}

\subsection{Weak Faraday Rotation}
\label{subsec:expansion}

The mean squared gradient of the MHD cubes and of the synthetic cube with the coherent source follows Equation~\eqref{eq:expansion}, with its coefficients computed from the same cube at zero wavelength, until the rotation is no longer small on the lines of sight that contribute. The supersonic cube leaves the expansion first, which we attribute to the intermittency of its rotation measure \citep{Melnichenka2026b}, which makes the rotation large on some lines of sight already at small $\eta$. The term in $\lambda^2$ is present in single maps and changes sign between observer positions. Reversing the magnetic field leaves the emissivity unchanged and reverses the RM density, and with the LOS perpendicular to the mean field both configurations are equally probable, so the cross term vanishes on average.

For a uniform source the gradient spectrum is the Faraday term alone (Figure~\ref{fig:weak}). Divided by $\eta^2$ the spectra at $\eta=0.1$, shown, and at $\eta=0.4$ coincide, as the growth with $\lambda^4$ requires, and they follow $K^2P_q/\sigma_\Phi^2$ computed from the prescribed spectrum. The slope is $-(1+m_\phi)$ over the whole range $0<m_\phi<2$, and for steep spectra it keeps steepening and does not level off at $K^{-2}$, as \LPM{} predict.

\subsection{Strong Turbulent Faraday Rotation}
\label{subsec:near}

Turbulent Faraday rotation turns the polarization angle by about one radian over the LOS distance
\begin{equation}
\ell=\big(\sqrt2\,\lambda^2\sigma_\phi\big)^{-1}
\label{eq:ell}
\end{equation}
(\LPM, Equation~(16), and $\mathcal L_{\sigma_\phi}$ of \LP, Equation~(39)), and the rotation is strong when $\ell$ is much smaller than $r_\phi$ and $L$. The normalized correlation of the polarization at two lines of sight separated by $R$, the coherence $\rho_P(R)$, follows from \LP{} for Gaussian fluctuations of the RM density. For a uniform source the two lines of sight remain coherent over the whole depth within the core radius $R_c$, and only over a depth comparable to the separation beyond the radius $R_a$, with
\begin{equation}
R_c=\left(\frac{\ell^2r_\phi^{\widetilde m_\phi}}{L}\right)^{\frac{1}{1+\widetilde m_\phi}}
\label{eq:Rc}
\end{equation}
and
\begin{equation}
R_a=r_\phi\left(\frac{\ell}{r_\phi}\right)^{\frac{2}{2+m_\phi}}
\label{eq:Ra}
\end{equation}
up to factors of order unity (\LPM, Equations~(21) and (22)), where $\widetilde m_\phi=\min(m_\phi,1)$. We write $K_c=1/R_c$ and $K_a=1/R_a$.

A pair of polarizations that is jointly Gaussian with zero mean and the coherence $\rho$ has the directional correlation
\begin{equation}
\mathcal F(\rho)=\frac{\pi}{4}\;{}_2F_1\!\left(\tfrac12,\tfrac12;2;|\rho|^2\right)\rho
\label{eq:F}
\end{equation}
(\LPM, Equation~(6)), where ${}_2F_1$ is the hypergeometric function. The pair observed through a turbulent Faraday-rotating volume is Gaussian only at given Faraday RMs along the two lines of sight, and the directional correlation $\xi_u(R)=\avg{u(\bm X)u^*(\bm X+\bm R)}$ is the average of Equation~\eqref{eq:F} over these depths. For a uniform source the randomness that remains at a given difference $\delta\Phi(z)$ of the two Faraday RMs is their mean $\Phi_s(z)$, and
\begin{equation}
\xi_u(R)=\avg{\mathcal F(\rho_\delta)}_{\delta\Phi}
\label{eq:cond}
\end{equation}
with the conditional coherence
\begin{equation}
\rho_\delta=\frac{\int\!\!\int dz\,dz'\,G(z,z')\,e^{i\lambda^2[\delta\Phi(z)+\delta\Phi(z')]}}{\int\!\!\int dz\,dz'\,G(z,z')\,e^{i\lambda^2[\delta\Phi(z)-\delta\Phi(z')]}}
\label{eq:rhodelta}
\end{equation}
(\LPM, Equation~(B33)), where $G(z,z')=\exp\{-2\lambda^4\avg{[\Phi_s(z)-\Phi_s(z')]^2}\}$. This form requires a volume much deeper than $r_\phi$. We compute $\rho_P$ from the prescribed spectrum of the RM density with the correlation of \LP, and average Equation~\eqref{eq:cond} over realizations of $\delta\Phi$ drawn from the same spectrum.

In a deep volume, where the difference of the Faraday RMs grows with depth as a random walk, two thirds of the decorrelation $\varepsilon=1-\rho_P^2$ of nearly coherent lines of sight is a rotation of one integrated polarization relative to the other, which does not change the interference of the emission from different depths, and
\begin{equation}
1-\xi_u=\frac{\varepsilon}{12}\left[\ln\frac{16}{\varepsilon}+C_B\right]
\label{eq:brown}
\end{equation}
with $C_B=3.79$ (\LPM, Equation~(14)), a leading coefficient three times smaller than that of the Gaussian pair, $1-\xi_u=(\varepsilon/4)[\ln(16/\varepsilon)-1]$.

Inside the core the decorrelation grows as $\varepsilon=(R/R_c)^{1+\widetilde m_\phi}$ for the uniform source, and for $m_\phi<1$ the gradient spectrum at $K\gg K_c$ and the smoothed variance are
\begin{equation}
P_{\nabla u}(K)\propto K^{-(1+m_\phi)}\left[\ln(KR_c)+{\rm const}\right]
\label{eq:coreP}
\end{equation}
and
\begin{equation}
\avg{|\nabla u_b|^2}\propto b^{m_\phi-1}\ln(R_c/b)
\label{eq:coreV}
\end{equation}
(\LPM, Equation~(38)), with amplitudes that follow from Equation~\eqref{eq:brown}. For $m_\phi>1$, where $\varepsilon\propto R^2$, the spectrum is $K^{-2}$ without a logarithm. Outside the core, where the coherence is low, the gradient spectrum for $m_\phi<1$ has the limiting form
\begin{equation}
P_{\nabla u}(K)\propto K^{1+m_\phi}
\label{eq:inner}
\end{equation}
for $K_a\ll K\ll K_c$ (\LPM, Table~2).

The mean squared gradient at a given resolution is set by the decorrelation of the directions of neighboring lines of sight, $1-\xi_u$, shown in Figure~\ref{fig:near} against the decorrelation of the polarization. The maps with randomized phases follow the Gaussian pair, Equation~\eqref{eq:F}. The simulated maps have the same coherence and lie below it, increasingly so toward full coherence, although the polarization at a single sky position is Gaussian in these volumes. A Gaussian distribution at each position does not make the pair Gaussian. The conditional average of Equation~\eqref{eq:cond} follows the simulated maps, except in the shallowest cube with a steep spectrum, which holds too few correlation lengths of the RM density for the Gaussian limit it assumes. Near coherence the maps approach Equation~\eqref{eq:brown}, because a relative rotation lowers the coherence of the polarization as much as random mixing would but changes the directions less.

In the cubes that resolve the core (Figure~\ref{fig:core}), the gradient spectrum rises through the low-coherence interval, peaks near $K_c$ and falls inside the core, and the full prediction, the coherence of \LP{} followed by the conditional average, follows it over all wavenumbers. Inside the core the spectrum and the smoothed variance are set by the near-coherence form, Equation~\eqref{eq:brown}, and the Gaussian pair with the same coherence overestimates both, by up to the factor of three between the leading coefficients. The shape is that of Equations~\eqref{eq:coreP} and \eqref{eq:coreV}. The logarithm makes the spectrum shallower than $K^{-5/3}$ for $m_\phi=2/3$, and for $m_\phi=3/2$ it is $K^{-2}$ without a logarithm, while the smoothed variance grows toward small beams as $b^{-1/3}$ times a logarithm and only logarithmically, respectively. Between $K_a$ and $K_c$ the spectrum does not reach the limiting slope of Equation~\eqref{eq:inner}, because a cube that resolves the core separates the two scales by only a decade, and the maps follow the full prediction there.

\subsection{Strong Mean Faraday Rotation}
\label{subsec:mean}

When the RM density is constant, the Faraday RM is the same on all lines of sight and the mean Faraday rotation turns the polarization by one radian over the LOS distance
\begin{equation}
\Lb=\big(\lambda^2|\bar\phi|\big)^{-1}
\label{eq:Lb}
\end{equation}
(\LPM, Equation~(16), and $\mathcal L_{\bar\phi}$ of \LP, Equation~(37)). The polarization map is then the Fourier mode of the emissivity along the line of sight with the wavenumber $2/\Lb$. For a Gaussian emissivity with an isotropic spectrum $k^{-3-m_i}$ the coherence is
\begin{equation}
\rho_P(R)=\frac{2^{1-\nu}}{\Gamma(\nu)}\left(\frac{2R}{\Lb}\right)^{\nu}\mathcal K_\nu\!\left(\frac{2R}{\Lb}\right)
\label{eq:meanrho}
\end{equation}
(\LPM, Equation~(30)), where $\nu=(1+m_i)/2$ and $\mathcal K_\nu$ is the modified Bessel function of the second kind. The observed pair is jointly Gaussian, so $\xi_u=\mathcal F(\rho_P)$, and the gradient spectrum and the smoothed variance depend on the wavelength only through $K\Lb$ and $b/\Lb$. For $0<m_i<1$ they approach
\begin{equation}
P_{\nabla u}(K)\propto K^{-(1+m_i)}\left[\ln(K\Lb)+c_0\right]
\label{eq:meanP}
\end{equation}
for $K\Lb\gg1$ and
\begin{equation}
\avg{|\nabla u_b|^2}\propto\Lb^{-(1+m_i)}\,b^{m_i-1}\left[\ln(\Lb/b)+C_b\right]
\label{eq:meanV}
\end{equation}
for $b\ll\Lb$, with $c_0=1.30$ and $C_b=-1.87$ for $m_i=2/3$, the constants of \LPM{} (Equations~(B48) and (B51)) written in the variables $K\Lb$ and $\Lb/b$. Both limits are approached slowly, because the corrections fall only as $(K\Lb)^{-(1-m_i)}$ and as a logarithm.

In the variables $K\Lb$ and $\Lb/b$ the two wavelengths fall on one curve (Figure~\ref{fig:mean}), so the peak of the gradient spectrum moves to larger wavenumbers as $\lambda^2$ and the variance at a given beam grows with wavelength. The transform of $\mathcal F(\rho_P)$ follows the measured spectrum and variance. The small excess at the longer wavelength is in the coherence of these maps, which sample few coherence scales, and disappears when the measured coherence is used. The emissivity is quadratic in Gaussian fields and not itself Gaussian, and the relation holds nevertheless. Over the range of the maps the asymptotic forms, even with the constants $c_0$ and $C_b$, lie above the measurements, which only the full transform describes.

\section{Discussion}
\label{sec:discussion}

The gradient spectrum, the smoothed variance, the coherence $\rho_P$ and the directional correlation $\xi_u$ are all obtained from one pair of Stokes maps at one wavelength, and the test of Figure~\ref{fig:near} needs no model. Comparing the decorrelation of the directions with that of a map with randomized phases shows whether neighboring lines of sight differ by a relative rotation. The regime itself has to be identified with more than one wavelength, since the Faraday term grows as $\lambda^4$ under weak Faraday rotation and the peak of the gradient spectrum moves as $\lambda^2$ under strong mean Faraday rotation and as $\lambda^{4/(1+\widetilde m_\phi)}$ under strong turbulent Faraday rotation.

The MHD cubes test the gradient predictions under weak Faraday rotation only. Under strong turbulent Faraday rotation the rotation measure of MHD turbulence is intermittent, and the near-coherence relation of Equation~\eqref{eq:brown}, derived for a random walk of the Faraday RMs and a uniform source, remains to be tested with a turbulent source in simulations that resolve the core. With the LOS along the mean field the unrotated polarization has no preferred direction, so the expansion of Equation~\eqref{eq:expansion} does not apply. Apart from the noise test of Section~\ref{sec:compare}, our maps are free of noise, and the beam is applied to the direction, while in observations the finite angular resolution acts on $Q$ and $U$ before the normalization, and noise dominates the direction where the polarized intensity is low. Both effects are expected to weigh more on the gradient than on the directional spectrum \citepalias{Lazarian2026} and have to be modeled before the method is applied to surveys \citep{Heald2020}. In observations, gradient maps have been read as Faraday structure only where the polarized structure is not intrinsic to the source \citep{Sun2014}, and gradients need fewer frequencies than Faraday tomography \citep{Ho2019}. Here emission and rotation are not separable, and the comparison indicates which single-frequency statistic still isolates the rotation.

\section{Summary}
\label{sec:summary}

We compared the normalized tangential polarization gradient with four other polarization measures on synthetic observations of \textsc{AthenaK} MHD simulations and of synthetic data cubes with spatially coincident synchrotron emission and Faraday rotation, and tested the predictions of \LPM{} for the gradient.

\begin{enumerate}
\item The normalized gradient and the directional spectrum, which are one statistic, follow the rotation measure spectrum at least as well as the spectrum of the complex polarization. They reach the Faraday slope at the same rotation strength, and when the polarized intensity of the source fluctuates their slope is closer to it at weak rotation.

\item The spectrum of the polarization derivative $\partial P/\partial\lambda^2$ has the Faraday slope from the shortest wavelengths, as \LP{} predicted, and is the most direct measure with noise-free maps at two wavelengths, but it does not reach the Faraday slope in our noise test.

\item The spectrum of the magnitude of the polarization gradient keeps its slope under rotation in the MHD cube, as \citet{ZhangLiu2025} found, and in neither cube keeps the Faraday slope, which the vector gradient carries in both.

\item Under strong turbulent Faraday rotation neighboring lines of sight are not a Gaussian pair, although the polarization at each of them is Gaussian, and most of their decorrelation is a relative rotation, which lowers the gradient power in the coherent core relative to Gaussian maps of the same coherence.

\item Under weak Faraday rotation the mean squared gradient follows its expansion in $\lambda^2$, and the Faraday term of the gradient spectrum grows as $\lambda^4$ with the slope $K^{-(1+m_\phi)}$ for shallow and steep spectra of the RM density, as \LPM{} predict.

\item Under strong mean Faraday rotation the gradient spectrum and the smoothed variance depend on the wavelength only through $K\Lb$ and $b/\Lb$ and follow the transform of the Gaussian-pair correlation, as \LPM{} predict.
\end{enumerate}

\section*{Acknowledgments}
A.L. acknowledges the support of NSF grant AST 2307840.

\begin{center}ORCID iDs\end{center}
\noindent A. Melnichenka\,\includegraphics[width=9pt]{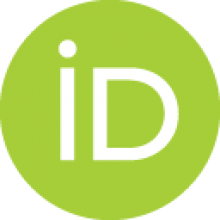} \href{https://orcid.org/0009-0004-1969-779X}{orcid.org/0009-0004-1969-779X}\\
A. Lazarian\,\includegraphics[width=9pt]{Orcid-ID.png} \href{https://orcid.org/0000-0002-7336-6674}{orcid.org/0000-0002-7336-6674}\\
D. Pogosyan\,\includegraphics[width=9pt]{Orcid-ID.png} \href{https://orcid.org/0000-0002-7998-6823}{orcid.org/0000-0002-7998-6823}

\newpage
% written by code/make_bib.py, do not edit by hand

\end{document}